\pdfoutput=1

\documentclass[conference]{IEEEtran}
\usepackage{cite}

\ifCLASSINFOpdf
  \usepackage[pdftex]{graphicx}
   \DeclareGraphicsExtensions{.pdf,.jpeg,.png}
\else
   \DeclareGraphicsExtensions{.eps}
\fi
\usepackage{amsmath}
\usepackage{multirow}

\usepackage{listings}

\usepackage{url}

\begin{document}
%
\title{OSDF: A Framework For Software Defined Network Programming}

\author{\IEEEauthorblockN{Douglas Comer\IEEEauthorrefmark{1}, Adib Rastegarnia \IEEEauthorrefmark{2}}
\IEEEauthorblockA{ Dept of Computer Science,
Purdue University, USA, IN, 47907\\
Email: \IEEEauthorrefmark{1}comer@cs.purdue.edu,
\IEEEauthorrefmark{2}arastega@purdue.edu,
}}


%


\maketitle

\begin{abstract}
Using SDN to configure and control a multi-site network involves writing code that handles
low-level details.  We describe preliminary work on a framework that takes a network description and set of policies as input, and handles all the details of deriving routes and installing flow rules in switches.  The paper describes key software components and reports preliminary results.
\end{abstract}


%
\IEEEpeerreviewmaketitle

\section{Introduction}
Software Defined Networking (SDN) is an emerging trend for the design of network management systems. SDN breaks vertical integration by decoupling the control plane from data plane, and provides flexibility that allows software to program the data plane hardware directly. The separation of the
control and data planes allows network switches to become simplistic forwarding devices, and allows
control logic to be implemented by a logically centralized controller called a Network Operating System
(NOS)\cite{Ref1}. In recent years, SDN has received attention from service providers, enterprises, and researchers.
However, a set of open problems remain unsolved, and must be addressed before the ultimate promise of SDN can be fulfilled.
In the current SDN paradigm, an external SDN controller communicates with management applications and with one or more network devices to configure and control the devices. Communication requires that the two communicating parties agree on an Application Program interfaces (API) to be used.
Current SDN controllers employ two APIs known as \emph{Northbound} and \emph{Southbound}. The Southbound API defines communication between the controller and a network device.
Early SDN work defined the OpenFlow \cite{Ref7} southbound API, and the OpenFlow protocol continues to dominate the southbound protocol space. OpenFlow allows a controller to update flow table rules, and to specify associated
actions to be performed for each of the flows that pass through a given network devices \cite{Ref2}. The current software-defined management architecture exhibits several weaknesses as follows \cite{Ref3}:
\begin{itemize}
\item \emph{Low-level interface}. Current southbound APIs control low-level details. For example,
although it provides operations agnostic to any vendor, OpenFlow only handles basic details. Proposed programming systems, such as Frenetic \cite{Ref6}, provide high-level abstractions
to control a network directly, but require a programmer to map abstractions to OpenFlow rules that associate actions with rules that match fields in a packet. To overcome the weaknesses, a set of new abstractions must be designed that allow programmers to focus on end-to-end application requirements instead of the low-level details that OpenFlow exposes.
Consequently, we are working to create abstractions that reflect end-to-end requirements.
\item \emph{Inadequate functionality}. A flow-based, match-action abstraction cannot specify complex
network functions, such as encryption/decryption functions used for security and deep packet inspection (DPI) used for malware detection. Because such functions require examination or transformation of the payload in a packet, header field matching is insufficient. To combat the weakness we propose to focus on abstractions that go beyond flow-based,
match-action rules and support flexible network functions.
\item \emph{Low-level programming interface}. Typical SDN controllers provide a simplistic REST-ful API which means a programmer must specify low-level the details for each flow, and
must parse JSON or XML formats to retrieve network topology, network parameters, and statistics. One of the critical pieces of our work arises from the design of a new interface that allows programmers to write network management application without worrying about the
low-level details of how the information is transported and stored, and without any need to
write code that parses data obtained from a device. In essence, we propose to find ways to remove the drudgery from network programming.
\end{itemize}
To overcome the weaknesses and achieve the goals outlined earlier, we designed a high-level framework for Software Defined Network Programming, and built and early prototype we call \emph{Open Software Defined Framework }(OSDF) that uses the Open Network Operating System (ONOS) \cite{Ref4,Ref5} as an SDN controller. The rest of the paper is organized as follows: the following section presents the overall design goals and Section \ref{architecture} gives and overview of the architecture.  Sections \ref{polices} and \ref{hlnp}, present the details of application based network polices and high level network operations that we support. Section \ref{results}, presents preliminary experimental results. We present the future work in Section \ref{fws} and  Section \ref{conclusions} concludes the paper. 
\section{An Overview Of The Framework Goals}
\label{second}
\subsection{Overall Design Goals}
\begin{enumerate}
\item Provide a high level interface that allows managers to express policies and, when needed, allows programmers to write network management applications without worrying about the low-level details of how the information is transported and stored, and without writing code that parses data obtained from a device.

\item Predefine a set of high level network services that can be invoked by management applications to configure a switch without knowing the details of the southbound API (e.g., OpenFlow or an alternative). 

\item Devise a system that can run management applications analogous to the
way a conventional operating system runs processes. Just as a conventional process uses services provided by its operating system, a network management application will use services provided by our system.

\item Design and develop a hybrid approach that allows programmers to specify network configurations both \emph{proactively}, by deriving configuration rules from high-level network policies, and \emph{reactively}, by modifying the configuration as flows and conditions change.

\end{enumerate}

\section{Framework Architecture}
\label{architecture}
\begin{figure}[ht]
\centering
\includegraphics[scale=0.27]{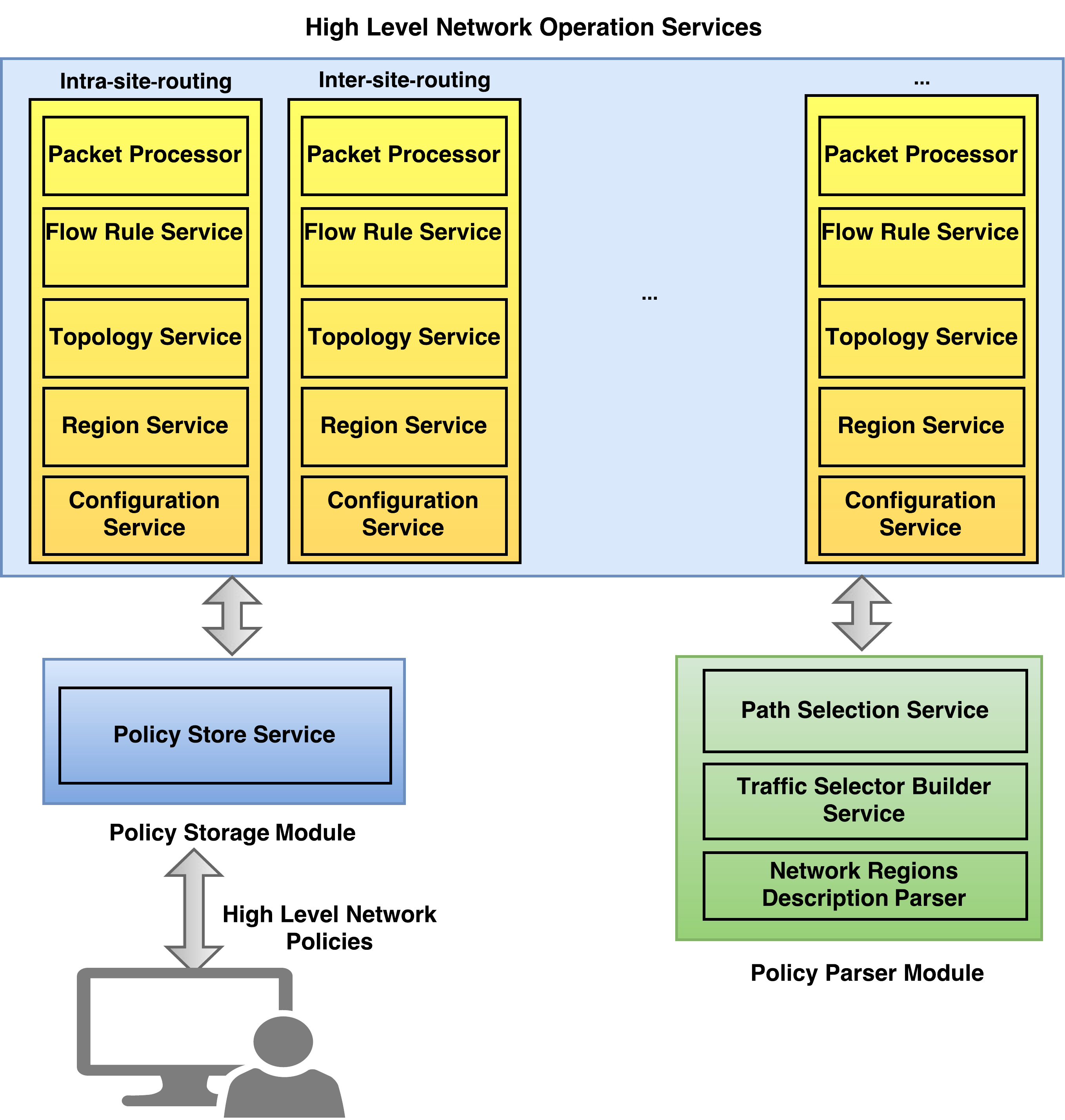}
\caption{OSDF Architecture}
\label{Fig1}
\end{figure}

Figure \ref{Fig1} illustrates the framework architecture, which consists of the key components described below.

\begin{itemize}
\item \emph{High level network operation services}: A set of abstract operations used to configure and monitor a network based on high level network polices that a network administrator provides. Each service reads the high level network polices from a database of currently active polices, parses them to generate network-wide forwarding rules for incoming flows, and finally, installs rules in appropriate network switches. Each operation service uses a hybrid approach in which the service can create basic rules proactively from requirements specified in the policies, and can reactively extract information from incoming packets and use the information to generate, install, and update rules.  Each network operation service includes the following subcomponents: 
\begin{itemize}
\item \emph{Packet Processor}: This component parses incoming packets that reach the controller, and extracts low level match fields such as the source and destination IP addresses, MAC addresses, ports, protocol number, and so on. The Packet Processor reactively processes incoming packets which offers an administrator the flexibility to defer configuration and hide all the low level details that are needed to configure and monitor the network. Packet processors parse the currently active policies that are stored in the policy storage, and use the policies to select a subset of match fields. Depending on type of application and high level requirements that an administrator specifies in a network policy, a subset of low level parameters will be selected by a \emph{Traffic Selector Builder Service} to generate a set of match fields for a flow. In addition, depending on the operation, a corresponding Packet Processor service chooses an appropriate action for each of the flow rules. For example, in intra-site routing, the flow rules associated with a given flow specify forwarding packets to appropriate outgoing port in each switch.  For inter-site routing, rules may specify rewriting the MAC address, encapsulation, or other action.
\item \emph{Flow Rule Service}:  The Flow Rule Service uses the set of match fields generated by the Traffic Selector Builder Service to generate and install flow rules in appropriate network devices. 

\item \emph{Topology Service}: We use the Topology Service to find and determine an appropriate path for incoming flows based on high level network policies. The topology information which we obtain from the Topology Service is used by the \emph{Path Selection Service} which is explained below.

\item \emph{Region Service}: A \emph{region} refers to a group of devices located in a common physical (i.e., geographical) or logical region. The Region Service provides information about devices inside a region.  The information can used by network regions description parser.   

\item \emph{Configuration Service}: The Configuration Service is responsible for items used in configuration, including both details of individual devices, their IDs and locations, the IP prefixes used, the mapping of prefixes to regions, and predefined items, such as default gateways.

\end{itemize}

\item \emph{Policy Store Module}: A module used to store and retrieve application-based network policies that an administrator enters to the system.  Abstract operation services use the Policy Store to read and parse policies and install flow rules accordingly. In addition, an administrator can update or delete network policies dynamically at runtime. 
\item \emph{Policy Parser Module}: The Policy Parser analyzes application-based policies and incoming flows, and derives a set of match fields that are then used to generate a set of flow table rules. The module includes of the following three subcomponents, which are invoked by abstract operation services: 
\begin{itemize}
\item \emph{Path Selection Service}: This service provides a set of pre-defined algorithms to choose among a set of existing paths between two end points (e.g., shortest path).  The module is extensible: a network programmer can specify additional path finding algorithms. 
\item \emph{Traffic Selector Builder}: The service generates a set of match fields based on polices and incoming packets.  The service is also extensible: a programmer can define new types of applications an specify combinations of packet match fields for each application.

\item \emph{Network Regions Parser}: This service uses information that the Region Service provides, and parses incoming flows and categorizes them based on the regions they span. 
\end{itemize}
\end{itemize}

\section{Application-Based Network Policies}
\label{polices}
An application-based network policy specifies high level requirements for a given application (type of packet).
The policies are used to configure and monitor network devices. We categorize all policies into two major categories: \emph{intra-domain policies} (inside a region or a site) and \emph{inter-domain policies} (among multiple regions or sites).  An application policy includes the following key items:
\begin{itemize}
\item \emph{Traffic Profile}: A traffic profile specifies high level characteristics and requirements for an application, such as an application name (e.g. VOIP), the transport protocol used (e.g TCP or UDP), and a traffic type (e.g. real time vs best effort). The system provides a set of pre-defined traffic profiles that support typical , such as web, video, and voice traffics but a network administrator can extend it by introducing new traffic profiles to support new types of applications. 
\item \emph{High Level Network Function}: Each policy is associated with a high level network function that we defined for configuring and monitoring of network devices such as intra-site-routing and inter-site-routing. Associating each policy with a network operation allows a Packet Processor to accommodate polices that are related to the function and ignore non-relevant functions.   

\item \emph{Partial Hosts And Devices Information}: An administrator has the flexibility to provide high-level information about devices and hosts (e.g a name for a host or network device).  The high-level information can be used by the Path Selection Service when it determines an end-to-end path between the source and destination for specific traffic. For example, if an administrator chooses to route  a specific type of traffic through a specific set of network devices, the administrator can specify the requirements in a network policy and allow Path Selection Service to choose the best possible path that meets the given requirements.  

\item \emph{Priority}: An administrator can assign a priority to a policy or allow the system to use a default.  Priorities become important when policies overlap.

\item \emph{Source And Destination Regions}: A policy can be defined for the interior of a region or to specify traffic routing among multiple regions. To achieve the goal, the system allows a manager to specify both source and destination regions for each policy.   

\end{itemize}

\section{High level Network Operations}
\label{hlnp}
In the initial prototype version of the framework we define a minimum set of high level network operations to support typical network configurations:

\begin{itemize}
\item \emph{Intra-Site-Route:} This abstract operation can be used to specify traffic routing within a specific region according to a network policy for the region.
\item \emph{Inter-Site-Route:} This abstract operation can be used to specify traffic routing among multiple regions according to the global policies.
\end{itemize}

\section{Experimental Results}
\label{results}
This section demonstrates how the high-level network operations described above provide support for high-level policies and can be used to configure the networks accordingly. In addition, we report the response time of a prototype when policies are added, removed and changed. 
\subsection{Framework Usage}
\begin{figure}[ht]
\centering
\includegraphics[scale=0.24]{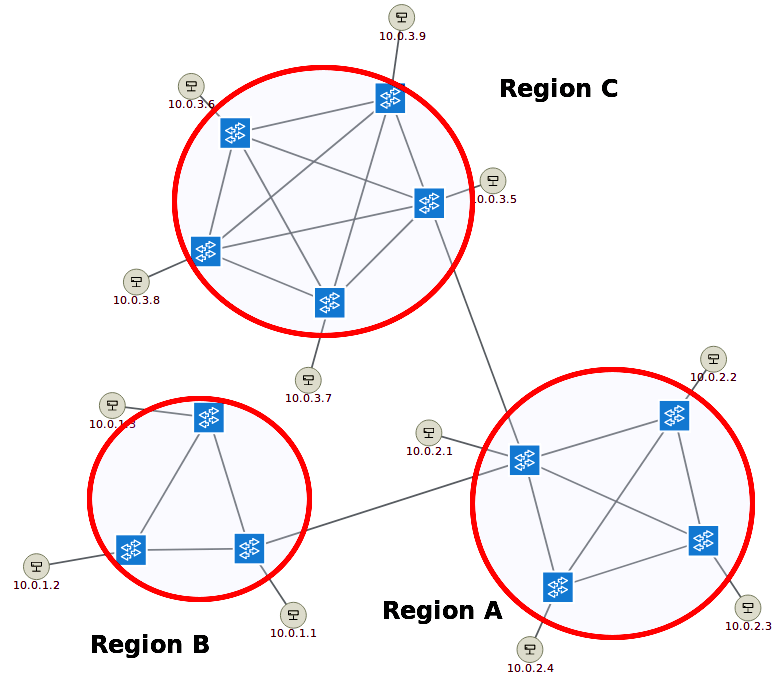}
\caption{An Example of Multiple Regions Network Topology}
\label{Fig2}
\end{figure}

We use the the network topology that Figure \ref{Fig2} illustrates to explain the framework usage.

The network has three sites that we call \emph{regions}, with each region assigned a unique IP prefix. Suppose we want to configure the network to support the following policies:

\begin{itemize}
\item Route Web traffic between region A and region B with priority 100.
\item Route Ping traffic between region B and C with a default priority.
\item Route Video traffic inside regions A and C with priority 300.
\end{itemize}

To implement the above, we provide a basic set of functions that can be executed from the controller CLI to create and enter policies. As we mentioned earlier, after inserting polices into the system, if we initiate traffic between two end-points inside a region or between two sites, the system must have translated the policies into a set of flow rules and installed the rules to enable endpoints to communicate with each other. 

\subsection{Experimental Setup}
To evaluate our framework, we used the Mininet network emulator \cite{Ref8} and ran experiments on a Linux host with a Intel CORE i7 processor and 8GB of RAM. 

\subsection{Performance Evaluation}
\begin{figure}[ht]
\centering
\includegraphics[scale=0.4]{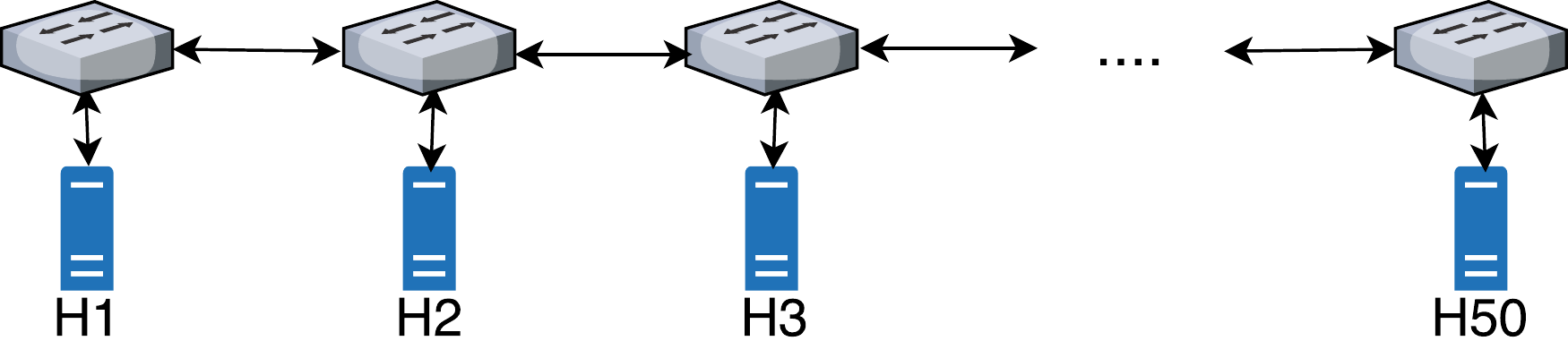}
\caption{Linear Network Topology}
\label{Fig3}
\end{figure}

To evaluate the efficiency of our framework, we measure the response time vs. path length for two scenarios including OSDF and ONOS reactive forwarding approaches in a linear network topology  illustrated in Figure \ref{Fig3}.

\begin{figure}[ht]
\centering
\includegraphics[scale=0.62]{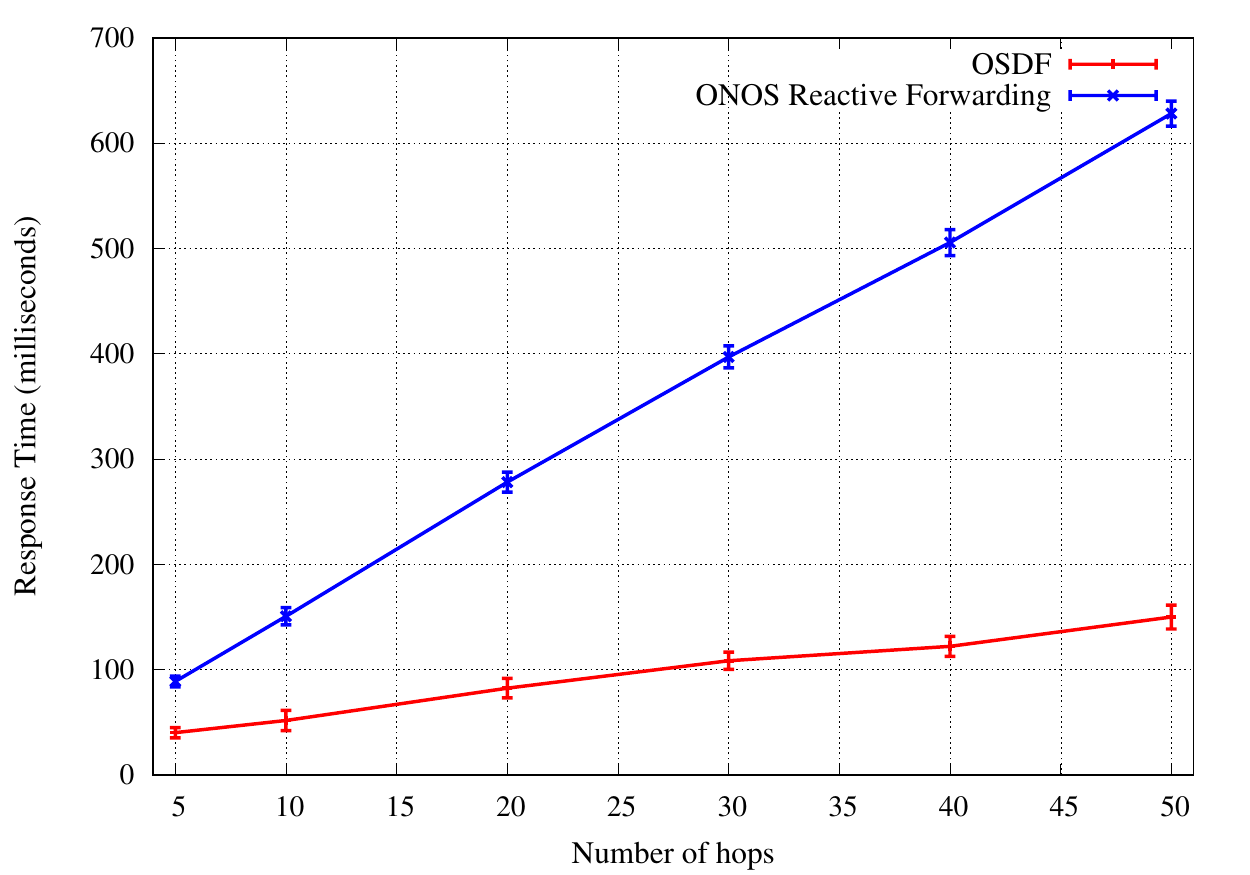}
\caption{Response Time vs Path Length (95\% of Confidence Interval for Mean)}
\label{Fig4}
\end{figure}

In OSDF, we define the response time as the amount of time which is needed to read and parse a network policy, generate flow rules , and install generated rules on the network devices. The same definition can be applied to the reactive forwarding approach excluding the time which is needed to read and parse a network policy.  As the results in Figure \ref{Fig4} show, the time required to process and install rules increases close to linearly as expected in both of the scenarios. The reason that our approach outperforms the reactive approach in ONOS, arises because we optimized the flow rule installation phase by installing all of the required flow rules for an end to end path when the controller receives the first \emph{PACKET\_IN} message. Consequently, we reduce the end to end response time by reducing the number of \emph{PACKET\_IN} messages that must be processed by the controller. By comparison, the ONOS reactive approach configures each switch separately by waiting until a \emph{PACKET\_IN} message arrives from a switch.       

\section{Future Work}
\label{fws}
We plan to complete the framework by introducing additional abstract operations and adding new features.  Our goals include:
\begin{itemize}
\item Design and develop new abstract operations to support some of the key functional units, such as an Intrusion Detection System (IDS), firewall, VPN, load balancer, and Network Address Translator (NAT).
\item Design and develop a subsystem to resolve policy conflicts before installing flow rules into the network devices. 
\item Test and evaluate the efficiency of the framework on a real network.
\item Provide a user friendly interface that allows an administrator to configure a network without any need to write scripts, except for unusual cases.
\end{itemize}

\section{Conclusion and Remarks}
\label{conclusions}
In this paper, we proposed a new policy-based framework for software defined network programming. The proposed framework follows on a hybrid approach that allows an administrator to specify network configuration requirements proactively using a high level policy language. The framework reactively generates the required flow rules, and installs them into each network device. One of the main features of the framework is its ability to hide all of the low level details that are used to configure a network as policies change.       





\bibliographystyle{IEEEtran}
\bibliography{ref}

\end{document}